\documentclass[11pt]{article}
\pdfoutput=1
\usepackage{verbatim}
\usepackage[centertags]{amsmath}
\usepackage[square, comma, sort&compress,numbers]{natbib}
\usepackage{array,multirow}
\numberwithin{equation}{section}
\usepackage{amssymb}
\usepackage{color}
\usepackage[normalem]{ulem}

\usepackage{epsfig}
\usepackage{graphicx}
\usepackage{bbold}

\usepackage{wrapfig}
\usepackage{float}
\usepackage{soul}
\usepackage{url}

\usepackage{tikz,pgf}
\usetikzlibrary{shapes}
\usetikzlibrary{calc}
\usetikzlibrary{decorations.pathmorphing}
\usetikzlibrary{decorations.pathreplacing,shapes.misc}
\usetikzlibrary{positioning}
\usetikzlibrary{arrows}
\usetikzlibrary{decorations.markings}
\usetikzlibrary{shadings}

\usetikzlibrary{intersections}

\newcommand{\mf}[1]{\mathfrak{#1}}
\def\bsp{\begin{split}}
\def\esp{\end{split}}
\def\osp{\mathfrak{osp}(1|2)}

\def\sideremark#1{\ifvmode\leavevmode\fi\vadjust{\vbox to0pt{\vss
 \hbox to 0pt{\hskip\hsize\hskip1em
 \vbox{\hsize3cm\tiny\raggedright\pretolerance10000
  \noindent #1\hfill}\hss}\vbox to8pt{\vfil}\vss}}}

\def\be{\begin{equation}}
\def\ee{\end{equation}}
\def\ba{\begin{array}}
\def\ea{\end{array}}

\advance\textheight\topskip
\renewcommand{\tilde}{\widetilde}
\renewcommand{\hat}{\widehat}

\newcommand{\binner}[2]{%
  {\langle}\kern-4.15pt{\langle}#1{,}\,#2{\rangle}\kern-4.15pt{\rangle}}

\newcommand{\ffrac}[2]{\raisebox{.5pt}%
  {\footnotesize$\displaystyle\frac{#1}{#2}$}\kern1pt}

\def\cF{\mathcal{F}}

\def\cP{\mathcal{P}}

\def\cV{\mathcal{V}}

\numberwithin{equation}{section} \makeatletter
\@addtoreset{equation}{section}

\def\be{\begin{equation}}
\def\ee{\end{equation}}
\def\ba{\begin{array}}
\def\ea{\end{array}}

\newdimen\tableauside\tableauside=1.0ex
\newdimen\tableaurule\tableaurule=0.4pt
\newdimen\tableaustep
\def\phantomhrule#1{\hbox{\vbox to0pt{\hrule height\tableaurule
width#1\vss}}}
\def\phantomvrule#1{\vbox{\hbox to0pt{\vrule width\tableaurule
height#1\hss}}}
\def\sqr{\vbox{%
  \phantomhrule\tableaustep

\hbox{\phantomvrule\tableaustep\kern\tableaustep\phantomvrule\tableaustep}%
  \hbox{\vbox{\phantomhrule\tableauside}\kern-\tableaurule}}}
\def\squares#1{\hbox{\count0=#1\noindent\loop\sqr
  \advance\count0 by-1 \ifnum\count0>0\repeat}}
\def\tableau#1{\vcenter{\offinterlineskip
  \tableaustep=\tableauside\advance\tableaustep by-\tableaurule
  \kern\normallineskip\hbox
    {\kern\normallineskip\vbox
      {\gettableau#1 0 }%
     \kern\normallineskip\kern\tableaurule}%
  \kern\normallineskip\kern\tableaurule}}
\def\gettableau#1 {\ifnum#1=0\let\next=\null\else
  \squares{#1}\let\next=\gettableau\fi\next}

\def\cF{\mathcal{F}}

\def\cP{\mathcal{P}}

\def\cV{\mathcal{V}}

\numberwithin{equation}{section} \makeatletter
\@addtoreset{equation}{section}

\def\be{\begin{equation}}
\def\ee{\end{equation}}
\def\ba{\begin{array}}
\def\ea{\end{array}}

\def\ba{\begin{array}}
\def\ea{\end{array}}

\def\sl2{\mathfrak{su}(1,1)}
\def\slC2{sl(2,\mathbb{C})}

\def\3j{3$j$}

\def\pref1{\varkappa_{_{j_p,j_1}}}
\def\a{\Omega}

\def\centerarc[#1](#2)(#3:#4:#5);%
    {
    \draw[#1]([shift=(#3:#5)]#2) arc (#3:#4:#5);
    }

\usepackage{jheppub}

\makeatletter
\def\@fpheader{\vspace{-.1cm}}
\makeatother

\title{Wilson lines construction of $\mathfrak{osp}(1|2)$ conformal blocks}

\author{Vladimir Belavin, J. Ramos Cabezas}

\affiliation{Physics Department, Ariel University, Ariel 40700, Israel.}

\emailAdd{vladimirbe@ariel.ac.il, juanra@ariel.ac.il}

\abstract{We study N=1 superconformal theory in the context of AdS/CFT correspondence in the large central charge limit using Chern-Simons formulation of $3d$ gravity. In this limit conformal dimensions of a subclass of so-called light primary superfields remain finite and are governed by $\mathfrak{osp}(1|2)$ subalgebra of N=1 super-Virasoro algebra. We describe the construction of $\mathfrak{osp}(1|2)$ conformal blocks in terms of Wilson lines of the Chern-Simons $3d$ gravity. We consider examples of two and three-point blocks on the sphere and one-point torus blocks of light superprimary fields, which belong to finite-dimensional representations of  $\mathfrak{osp}(1|2)$. We study the correlation function for lower and upper components of the primary $\mathfrak{osp}(1|2)$ doublets and show that the associated conformal blocks 
are obtained via  Wilson line construction in Chern-Simons theory.}

\begin{document}
\maketitle
\flushbottom
\section{Introduction}
  In recent works \cite{Hartman:2013mia, Fitzpatrick:2014vua, Caputa:2014eta, Hijano:2015rla, Fitzpatrick:2015zha, Alkalaev:2015wia, Hijano:2015qja, Hijano:2015zsa,  Alkalaev:2015lca, Alkalaev:2015fbw, Banerjee:2016qca, Gobeil:2018fzy, Hung:2018mcn, Alekseev:2019gkl}, in the context of the AdS/CFT correspondence, conformal blocks in the large $c$ limit have been identified with geodesic networks in the $AdS$ spacetime. An alternative description of conformal blocks via the Wilson lines of the Chern-Simons (CS) $3d$ gravity was considered in \cite{ deBoer:2013vca, Ammon:2013hba, deBoer:2014sna, Hegde:2015dqh, Melnikov:2016eun, Bhatta:2016hpz, Besken:2017fsj, Hikida:2017ehf, Hikida:2018eih, Hikida:2018dxe, Besken:2018zro, Bhatta:2018gjb, DHoker:2019clx, Castro:2018srf, Kraus:2018zrn, Hulik:2018dpl, Castro:2020smu, Chen:2020nlj}. In this work we use the latter approach in order to study super-Virasoro conformal blocks.
 
In the Virasoro case the four-point correlation function on the sphere of four primary fields $\varphi_i$ with conformal dimensions $h_i, \bar{h}_i$ is
\be \label{ni1}
\langle \varphi_1(z_1,\bar{z}_1)...\varphi_4(z_4, \bar{z}_4)   \rangle=\sum_{h,\bar{h} }C_{12h}C_{h34}|w(h, h_i,z_i)|^2,
\ee
where $C_{12h}$, $C_{h34}$ are the structure constants, $(z_i, \bar{z}_i)$ are coordinates on the complex plane and $w(h_s, h_i,z_i)$ is the  four-point holomorphic conformal block\footnote{For the holographic description of higher point conformal blocks see \cite{Hulik:2016ifr, Rosenhaus:2018zqn, Alkalaev:2018nik, Fortin:2019zkm, Parikh:2019ygo, Jepsen:2019svc, Anous:2020vtw}.}.

Similarly, the correlation functions are defined on the torus, for example, the one-point correlation function of a primary field $\varphi_1$ on the torus is \footnote{Throughout this paper we omit factor $(q\bar{q})^{-\frac{c}{24}}$ which can be easily restored.} 
\be \label{ni2}
\langle \varphi_1(z_1, \bar{z}_1)  \rangle= \text{\Large{Tr}}_h\left(     q^{\mathbf{L}_0} \bar{q}^{\bar{\mathbf{L}}_0}\varphi_1(z_1, \bar{z}_1)    \right)=\sum_h C_{h h_1 h} |\cF(h, h_1, q)|^2,
\ee
where $Tr_h$ is the trace taken over descendent states associated with the intermediate primary field $\varphi_h$,
$q$ is the elliptic parameter of the torus $q=e^{2\pi i \tau }$ and $\mathbf{L}_0$ is the Virasoro generator, $\mathbf{L}_0| h \rangle=h| h \rangle$. Here $\cF(h, h_1, q)$ is the one-point holomorphic torus conformal block (for details see \cite{DiFrancesco:1997nk, Belavin:1984vu,  Ribault:2014hia, Hadasz:2009db})\footnote{For AdS/CFT correspondence in thermal AdS, relevant for the torus topology, see \cite{Gobeil:2018fzy, Alkalaev:2016ptm, Kraus:2017ezw, Alkalaev:2017bzx, RamosCabezas:2020mew}.}. 

In \cite{Bhatta:2016hpz, Bhatta:2018gjb, Besken:2016ooo, Fitzpatrick:2016mtp} the dual description of $\mf{sl}(2)$ and $\mf{sl}(3)$ conformal blocks (which are obtained from the Virasoro and $W_3$ conformal blocks respectively in the large central charge limit for light primary fields\footnote{The conformal dimension $h$ of a \textit{light primary field} (resp. \textit{heavy primary field}) $\varphi_h$ by definition behaves like $h=\mathcal{O}(c^0)$ (resp. $h=\mathcal{O}(c^1)$) in the limit $c\rightarrow \infty$. For the holographic description of conformal blocks in the context of Virasoro heavy-light fields see \cite{Belavin:2017atm, Kusuki:2018wcv, Kusuki:2018nms, Hijano:2018nhq, Anous:2019yku, Alkalaev:2019zhs, Chen:2019hdv, Alkalaev:2020kxz}.}) on the sphere in the context of CS theory was found. It was shown that $\mf{sl}(2)$ 2-point, 3-point and 4-point conformal blocks on the sphere can be represented as follows
\be \label{ni3}
\langle  s | \prod_{i=1}^ne^{\int_{z_i}^{z_b} J_1^{(i)}dz}|hw \rangle_i, 
\ee
where $n=2,3,4$ give 2-point, 3-point and 4-point holomorphic conformal blocks respectively. Here $J_1^{(i)}dz$ is the $\mf{sl}(2)$ CS connection on the sphere acting in spin-$j_i$ representation, $| hw \rangle_i$  is the highest weight state of the representation with spin $j_i=-h_i$, and $ z_b$ is an arbitrary point in the bulk where the Wilson line operators $Pe^{\int_{z_i}^{z_b}J_1^{(i)}dz}$ are joined together. The state $| s \rangle$  is a singlet state which belongs to the tensor product of the representations $j_i$ ($i=1,2,...,n$). The relation between conformal blocks and the expression (\ref{ni3}) involves two basic ingredients: the singlet state $| s \rangle$ representing a specific channel of OPE and the boundary state $| hw \rangle_i$ associated with the boundary primary field $\varphi_i$ 

\be \label{ni4}
|hw \rangle_i \rightarrow \varphi_i(z_i).
\ee

In \cite{Kraus:2017ezw} (see also \cite{Alkalaev:2020yvq}) it was proposed that $\mathfrak{sl}(2)$ one-point torus conformal block $\cF(-j, -j_1,q)$, where $-j$ and $-j_1$ are respectively the intermediate and external conformal dimensions\footnote{In the dual description $j$ and $j_1$ are the spins of finite-dimensional representations of $\mf{sl}(2)$ CS gauge group.}, can be computed by a prescription similar to (\ref{ni3}) by introducing the following modifications: Wilson line operators at the point $z_b$ are joined using the intertwining operator and two boundary points are identified modulo $2\pi \tau$, and the trace is taken over corresponding representation. Thus the proposal reads 
\be \label{prop1}
Tr_{j}\Bigg{(}W_{j    }[z_b, z_b+2\pi \tau]I_{j;j, j_1} \Bigg{)}\otimes W_{j_1}[z_b, z_1]|lw \rangle_1 = \cF(-j, -j_1,q).
\ee
Here $Tr_{j}$ is taken over the representation with  spin $j$, $z_1$ is the point on the boundary of the solid torus which is a geometric representation of the thermal AdS$_3$, $z_b$ is an arbitrary point in the bulk of AdS$_{3}$, $|lw \rangle_1$ is the lowest weight state of the representation with spin $j_1$ and $I_{j;j,j_1}$ is the intertwining operator associated with the representations $j$ and $j_1$. The factors $W_a[x,y]$ (for $a=j,j_1$; $x=z_b$ and $y= z_b+2\pi \tau,z_1 $) denote the Wilson line operators
\be \label{prop2}
W_{a}[x,y]=\cP\exp \bigg(-\int_{x}^{y} \a \bigg) = \exp \left( (x-y) (   J_1+\frac{1}{4} J_{-1}  ) \right),
\ee
where $\Omega$ is the flat connection
\be \label{prop3}
\Omega=\bigg(J_1+\frac{1}{4} J_{-1}\bigg) dz,
\ee
where $J_{1}$ and $J_{-1}$ are the lowering and raising operators of $\mf{sl}(2)$ in the representation $a$. 

The purpose of this work is to investigate the generalization of this construction on the $\mf{osp}(1|2)$ algebra. As explained in \ref{scft}, this algebra is relevant in the large $c$ limit of the  $N=1$ super CFT. We are interested in studying the correlation functions of superprimary fields\footnote{See \cite{Bershadsky:1985dq, Friedan:1984rv, friedan1986notes, Dorrzapf:1997rx, Alvarez-Gaume:1991nvs, Qiu:1986if, Eichenherr:1985cx} for details of $N=1$ super CFT. For the AdS/CFT in the context of superfields see also \cite{Chen:2016cms, Hikida:2018eih}.} in the Neveu-Schwarz (NS) sector   
   
\be \label{ni5}
\Phi_i(z,\theta, \bar{z}, \bar{\theta})=\varphi_i(z, \bar{z})+\theta \psi_i (z, \bar{z})+\bar{\theta} \bar{\psi}_i(z, \bar{z})+ \bar{\theta} \theta\tilde{\varphi}_i(z, \bar{z}),
\ee
where $(z, \bar{z})$ are holomorphic and antiholomorphic coordinates, and $(\theta, \bar{\theta})$ are Grassmann variables, $\varphi_i$ is a primary field with conformal dimensions $(h_i, \bar{h}_i)$, $\psi_i = [\mathbf{G}_{-\frac{1}{2}}, \varphi_i]$, $\bar{\psi}_i = [\mathbf{\bar{G}}_{-\frac{1}{2}}, \varphi_i]$, $ \tilde{\varphi}_i= \{  \mathbf{G}_{-\frac{1}{2}}, \bar{\psi}_i  \} $. We call $(\varphi_i, \psi_i, \bar{\psi}_i,  \tilde{\varphi}_i )$ \textit{components of the superfield} $\Phi_i$. We are interested in the correlation functions of light superfields in the spherical and toroidal topologies. The conformal dimensions of degenerate superfields are given by the Kac formula \cite{Friedan:1984rv}, in the large $c$ limit these conformal dimensions become
\be \label{ni5-1}
h_{1,m}= -\frac{m-1}{4}+\frac{1-m^2}{8c}+\mathcal{O}(1/c^2),
\ee
thus light degenerate NS superfields in the limit $c\rightarrow \infty$ have conformal dimensions $-j=\frac{1-m}{4}$, where $m$ is odd, therefore $j$ can take integer or half-integer values. Below, $j$ will be identified with the superspin  of a finite-dimensional representation of $\mf{osp}(1|2)$.

On the sphere, we are interested in studying the Wilson lines formulation for the correlation functions of light superfields. By using the expansion (\ref{ni5}) we can express the correlation function 
\be \label{ni6}
\langle   \Phi_1(z_1, \theta_1, \bar{z}_1, \bar{\theta}_1) ...\Phi_n(z_n, \theta_n, \bar{z}_n, \bar{\theta}_n) \rangle 
\ee
in terms of components, thus (\ref{ni6}) contains $4^n$ terms, we call each of these terms \textit{components} of the correlation function. We will concentrate on the components to which $\varphi_i$ and $\psi_i$ contribute, and we will investigate only their holomorphic dependence (due to the factorization of the symmetry into the holomorphic and antiholomorphic sectors, the consideration of the antiholomorphic dependence and of the components $\bar{\psi}_i$, $\tilde{\varphi}_i$ is identical, see \cite{Belavin:1984vu}). To this end it is convenient to formally set $\bar{\theta}_i=0$, then we get
\be   \label{ni7}
\langle   \Phi_1 ...\Phi_n \rangle= \langle  \varphi_1...\varphi_n \rangle+...+\theta_1 ...\theta_n \langle  \psi_1...\psi_n  \rangle.
\ee

The correlation functions on the torus can be also expressed in terms of the components of the rhs of (\ref{ni5}). We restrict ourselves to considering the contributions of $\varphi_i$ and $\psi_i$ in the one-point correlation function. By setting $\bar{\theta}_1=0$ we obtain
\be \label{ni8}
\begin{split}
\text{\Large{Tr}}_h\left[  q^{\mathbf{L}_0} \bar{q}^{\mathbf{\bar{L}}_0}  \Phi_1(z_1, \theta_1, \bar{z}_1, \bar{\theta}_1)     \right]=  \text{\Large{Tr}}_h\left[ q^{\mathbf{L}_0} \bar{q}^{\mathbf{\bar{L}}_0}  \varphi_1(z_1,\bar{z}_1) \right] + \theta_1 \text{\Large{Tr}}_h\left[ q^{\mathbf{L}_0} \bar{q}^{\mathbf{\bar{L}}_0}   \psi_1(z_1,\bar{z}_1) \right],
\end{split}
\ee
where $\mathbf{L}_0$ is the generator of Virasoro subalgebra of Neveu-Schwarz algebra. In the large $c$ limit, for light fields the trace $Tr_h $ in (\ref{ni8}) is reduced to the $\mf{osp}(1|2)$ subalgebra. We will see that this trace splits into the even and odd parts
\be \label{ni10}
\text{\Large{Tr}}_h[...]= \text{\Large{Tr}}_h^e[...] +\text{\Large{Tr}}_h^o[...],
\ee
where $Tr_h^e$ is the trace over the states with weights $h+n$, $n$ belongs to the set of nonnegative integers, $Tr_h^o$ is the trace over the states with weights $h+n+1/2$. 

\hfill\break
Below, we state the result of this work.

On the sphere, in the case of the two-point and three-point correlation functions, the conformal block of each component can be computed by means of equation (\ref{ni3}) with the following prescription: $| s   \rangle$ is a singlet which belongs to the tensor product of superspin-$j_i$ representations of $\mf{osp}(1|2)$; $J_1^{(i)}$ has to be replaced by $L_1^{(i)}$-the lowering generator of the CS gauge algebra $\mf{osp}(1|2)$ acting in the representation $j_i$; We will have two sorts of states, the highest weight states $| hw \rangle _i$ with weights $j_i$ and states with weights $j_i-1/2$ which we denote by $| hw-1/2 \rangle_i$. For a component that contains the fields $(\varphi_i, \psi_i)$, each field $\varphi_i$ in this component will correspond to a state $|hw \rangle _i$ and each field $\psi_i$ will correspond to a state $| hw-1/2 \rangle_i$ in (\ref{ni3}), thus schematically we have
\be \label{ni11}
\begin{split}
& |hw \rangle_i \rightarrow \varphi_i(z_i), \\ &
|hw -1/2\rangle_i \rightarrow \psi_i(z_i).
\end{split}
\ee

In the case of the one-point correlation function on the torus, we show that the conformal block of each component of the correlation function (\ref{ni8}) can be computed by means of the lhs of equation (\ref{prop1}) using following modifications: $Tr_j$ is the trace taken over the states of a $\mf{osp}(1|2)$ finite-dimensional representation with superspin $j=-h$; $j_1$ is the superspin of a $\mf{osp}(1|2)$ finite-dimensional representation (the conformal dimension $h_1$ = $-j_1$); $I_{j; j,j_1}$ is the $\mf{osp}(1|2)$ intertwining operator (this operator can be expressed in terms of the $\mf{osp}(1|2)$ Clebsch-Gordan coefficients); $W_a[z_b, z]$ (for $a=j,j_1$; $z= z_b+2\pi \tau, z_1 $) are the Wilson line operators only this time ($J_1, J_{-1}$) have to be replaced by ($L_1, L_{-1}$)-the generators of the $\mf{osp}(1|2)$ gauge algebra; Similarly to the sphere, on the lhs of (\ref{prop1})  we will have two sorts of states, the lowest weight states $| lw\rangle_{1}$ with weight $-j_i$ and the states with weight $-j_1+1/2$ which we denote by $| lw+1/2 \rangle_1$. For a component that contains the field $\varphi_1$, the field $\varphi_1$ in this component will correspond to the state $| lw\rangle_{1}$ and for a component that contains the field $\psi_1$, the field $\psi_1$ will correspond to the state $| lw+1/2 \rangle_1$ on the lhs of (\ref{prop1}), thus schematically we have
\be \label{ni12}
\begin{split}
& |lw \rangle_i \rightarrow \varphi_i(z_i), \\ &
|lw+1/2\rangle_i \rightarrow \psi_i(z_i).
\end{split}
\ee

The outline of the paper is the following: In section \ref{osptheory} we give a brief framework of the $\mf{osp}(1|2)$ CFT and $\mf{osp}(1|2)$ finite-dimensional representation theory. In section \ref{confb} we present the Wilson lines formulation of $\mf{osp}(1|2)$ conformal blocks in the spherical and toroidal topology. In section \ref{conclusion} we give our conclusions and comment on further developments of the Wilson lines formulation of $\mf{osp}(1|2)$ conformal blocks. Appendixes \ref{scft} and \ref{sl2b} describe the notations of the Neveu-Schwarz algebra and the explicit form of $\mf{sl}(2) $ one-point torus block respectively. 
\section{$\mf{osp}(1|2)$ conformal field theory} \label{osptheory}
The Neveu-Schwarz algebra (also termed $N=1$ NS super-Virasoro algebra) contains the $\mf{osp}(1|2)$ subalgebra. We are interested in studying the limit $c\rightarrow \infty$ of $N=1$  NS super-Virasoro algebra, in this limit, see appendix \ref{scft}, we need to keep only the generators of the $\mf{osp}(1|2)$ in order to have finite inner products of states (similar to the Virasoro and $W_3$ algebras in the large $c$ limit, which reduce to $sl(2)$ and $sl(3)$ algebras respectively). In this sense, in the large $c$ limit, the $N=1$  NS super-Virasoro algebra reduces to the  $\mf{osp}(1|2)$  algebra\footnote{For a review of $\mf{osp}(1|2)$ CFT see \cite{Alvarez-Gaume:1991nvs, Ennes:1997vt, Gotz:2005jz}.}. In the $N=1$ super CFT we have conformal superfields $\Phi_i$ (\ref{ni5}) and in the sequel we restrict ourselves to studying the $(\varphi_i, \psi_i)$ components and omit the antiholomorphic contribution, thus for simplicity, we will write superfields as follows  

\be \label{repth1}
\Phi_i(z,\theta)=\varphi_i(z)+\theta \psi_i(z).
\ee 

The $\mf{osp}(1|2)$ algebra has three even generators $\mathbf{L}_{\pm1,0}$ and two odd generators $\mathbf{G}_{\pm \frac{1}{2}}$, we have the following commutation and anticommutation relations
\be \label{repth2}
[ \mathbf{L}_m, \mathbf{L}_n]=(m-n)\mathbf{L}_{m+n}, \qquad [\mathbf{L}_n, \mathbf{G}_{\pm \frac{1}{2}}]=\left(   \frac{n}{2}\mp \frac{1}{2}   \right)\mathbf{G}_{n \pm \frac{1}{2}},\qquad \{\mathbf{G}_r,\mathbf{G}_s\}=2\mathbf{L}_{r+s},
\ee  
where $[,], \{,\}$ are the commutator and anticommutator respectively. $m,n=0,\pm1$. $r,s=\pm \frac{1}{2}$. It is assumed the conjugation rules $\mathbf{L}_{-m}=(\mathbf{L}_m)^{\dagger}, \mathbf{G}_{-s}=(\mathbf{G}_s)^{\dagger}$. We see that generators $\mathbf{L}_m$ form $\mf{sl}(2)$ subalgebra. 

The states in the representation $\mathcal{V}_{h_i}$ are obtained by applying the generators $(\mathbf{G}_{-\frac{1}{2}}, \mathbf{L}_{-1})$ to the hwv $|h_i \rangle$ associated with the field $\Phi_i$, thus any state can be written as follows 
\be \label{repth4}
| M, h_i \rangle=\mathbf{L}_{-1}^m\mathbf{G}_{-\frac{1}{2}}^k|h_i \rangle, \qquad M=\left( m,k \right) , \quad m\in \mathbb{N}_0,  \quad k=0,1,
\ee 
where $\mathbb{N}_0$ is the set of nonnegative integers. Generators $(\mathbf{L}_1, \mathbf{G}_{+\frac{1}{2}})$ annihilate the hwv and 

\be \label{repth5}
\mathbf{L}_0 | h_i \rangle=h_i | h_i \rangle.
\ee
The supermodule $\mathcal{V}_{h_i}$ can be written as a direct sum of two subspaces
\be \label{repth6}
\mathcal{V}_{h_i}= V_{h_i}\oplus  V_{h_i+\frac{1}{2}},
\ee
where $(V_{h_i}, V_{h_i+\frac{1}{2}})$ are two $\mf{sl}(2)$ Verma modules associated with the hwv $|h_i \rangle, | h_i +\frac{1}{2} \rangle$ respectively, where $ | h_i +\frac{1}{2} \rangle=\mathbf{G}_{-\frac{1}{2}}|h_i \rangle $. 
\subsection{Spherical and toroidal conformal blocks}
In section \ref{sph} we will show how the conformal blocks (CB) of the components of the two-point and three-point correlation functions  are obtained in CS theory. 
These two examples clarify how the idea proposed in \cite{Bhatta:2016hpz, Bhatta:2018gjb, Besken:2016ooo, Fitzpatrick:2016mtp} generalizes to the $\mf{osp}(1|2)$ algebra and provide evidence that any superconformal block could be obtained in CS theory with $\mf{osp}(1|2)$ gauge symmetry. The CBs under consideration are presented.

\textbf{The two-point correlation function} of superfields $(\Phi_1, \Phi_2)$ ($h_1=h_2=h$) is given by
\be
 \label{cor1}
 \langle \Phi_1(z_1, \theta_1)\Phi_2(z_2, \theta_2) \rangle =  \frac{C_{12}}{(z_1-z_2-\theta_1 \theta_2)^{2h}}=C_{12}\left( \frac{1}{(z_1-z_2)^{2h}}+ \frac{2h \theta_1 \theta_2}{(z_1-z_2)^{2h+1}}    \right).\\
\ee
Using (\ref{repth1}) we have

\be \label{cor2}
 \langle \Phi_1(z_1, \theta_1)\Phi_2(z_2, \theta_2) \rangle  =  \langle \varphi_1(z_1)\varphi_2(z_2) \rangle+\theta_1\theta_2 \langle \psi_1(z_1)\psi_2(z_2) \rangle,\quad 
 \ee
where
\be \label{cor3}
\begin{split}
& \langle \varphi_1(z_1)\varphi_2(z_2) \rangle  \propto \frac{1}{(z_1-z_2)^{2h}},  \\ &\langle \psi_1(z_1)\psi_2(z_2) \rangle \propto   \frac{1}{(z_1-z_2)^{2h+1}}.
\end{split}
\ee

\textbf{The three-point correlation  function} of superfields $(\Phi_1, \Phi_2, \Phi_3)$ is given by

\be \label{cor4}
\langle  \Phi_1(z_1, \theta_1)\Phi_2(z_2, \theta_2) \Phi_3(z_3, \theta_3) \rangle = \frac{1}{Z_{12}^{\gamma_{123}}Z_{13}^{\gamma_{132}}Z_{23}^{\gamma_{231}}} \left(   C_{123}+\eta \tilde{C}_{123}    \right),
\ee
where
\be \label {cor5}
\begin{split}
Z_{ij}= (z_i-z_j-\theta_i \theta_j), \quad \gamma_{ijk}= h_j+h_i-h_k, \quad \\ \eta = (Z_{12}Z_{13}Z_{23})^{-1/2}(Z_{23}\theta_1+Z_{31}\theta_2+Z_{12}\theta_3+\theta_1 \theta_2 \theta_3).
\end{split}
\ee
Similarly, using (\ref{repth1}) and expanding the rhs of (\ref{cor4}), we have

\be \label{cor6}
\Big{\langle}   \left(\varphi_1(z_1)+\theta_1 \psi_1(z_1)   \right)    \left(  \varphi_2(z_2)+\theta_2 \psi_2(z_2)  \right)    \left(      \varphi_3(z_3)+\theta_3 \psi_3(z_3)  \right)    \Big{\rangle}  = \Gamma_0 +\Gamma_1,
\ee
where

\be \label{cor7}
\begin{split}
&\Gamma_0= \frac{C_{123}}{(z_1-z_2)^{\gamma_{123}}(z_1-z_3)^{\gamma_{132}}(z_2-z_3)^{\gamma_{231}}}\left( 1+\gamma_{231}\frac{\theta_2 \theta_3}{(z_2-z_3)} +\gamma_{132}\frac{\theta_1 \theta_3}{(z_1-z_3)}+\gamma_{123}\frac{\theta_1 \theta_2}{(z_1-z_2)}  \right),  \\ & \\
& \Gamma_1 =  \tilde{C}_{123} \frac{    (z_2-z_3)\theta_1+(z_3-z_1)\theta_2+(z_1-z_2)\theta_3+(-1/2+h_1+h_2+h_3)\theta_1 \theta_2 \theta_3   }{(z_1-z_2)^{\gamma_{123}+1/2}(z_1-z_3)^{\gamma_{132}+1/2}(z_2-z_3)^{\gamma_{231}+1/2}} .
\end{split}
\ee

\textbf{The one-point correlation function on the torus} is given by equation (\ref{ni8}). Each term of the rhs of (\ref{ni8}) has been found in \cite{Alkalaev:2018qaz} and look as follows

\be  \label{cor8}  
 \text{\Large{Tr}}_h\left[ q^{\mathbf{L}_0}  \varphi_1(z_1) \right]= C_{h h_1h}B_0(h, h_1, q) ,
 \ee
 
\be    \label{cor8-1}
 \text{\Large{Tr}}_h\left[ q^{\mathbf{L}_0}  \psi_1(z_1) \right]= C_{h h_1+\frac{1}{2} h} B_1(h, h_1, q).
\ee
For our purposes it is convenient to keep structure constants $C_{h h_1h}$ and $C_{h h_1+\frac{1}{2} h}$ in the definition of CBs. $B_0(h,h_1,q)$ and $B_1(h,h_1,q)$ are \textit{the lower} and \textit{upper torus superblocks}, given by
\be  \label{cor9}
\begin{split}
B_0(h, h_1, q)= & \frac{q^h}{(1-q)^{h_1}}{}_{2}F_1(2h-h_1, 1-h_1, 2h|q)-\\ &-\frac{2h-h_1}{2h}\frac{q^{h+1/2}}{(1-q)^{h_1}}{}_2F_1(2h-h_1+1, 1-h_1, 2h+1|q),
\end{split}
\ee

\be  \label{cor10}
\begin{split}
B_1(h, h_1, q)= & \frac{q^{h}}{(1-q)^{h_1+\frac{1}{2}}}{}_{2}F_1(2h-h_1-\frac{1}{2}, -h_1+\frac{1}{2}, 2h|q)- \\ &-\frac{2h+h_1-\frac{1}{2}}{2h}\frac{q^{h+1/2}}{(1-q)^{h_1+\frac{1}{2}}}{}_2F_1(2h-h_1+\frac{1}{2}, -h_1+\frac{1}{2}, 2h+1|q),
\end{split}
\ee
where ${}_2F_1(a,b,c|z)$ is the hypergeometric function. We notice that (\ref{cor9}, \ref{cor10}) have poles when $h$ takes  negative integer or half-integer values, at these values we need to truncate the series expansion of BHS in (\ref{cor9}), (\ref{cor10}) according to the standard null-vector condition for degenerate fields.
\subsection{Finite-dimensional representations}
In this section we recall some facts of $\mf{osp}(1|2)$ finite-dimensional representation theory which are relevant for our discussion in the subsequent sections. 

A finite-dimensional representation of $\mf{osp}(1|2)$ is labelled by a nonnegative integer or half-integer \textit{superspin} $j$ and \textit{parity} $\lambda$ ($\lambda=0$ if $j$ is integer, or $\lambda=1$ if $j$ is half-integer), we call these representations \textit{superspin-$j$ representations} of $\mf{osp}(1|2)$, each representation decomposes into two normal subspaces as follows
 
\be \label{repth7}
\cV_{j} =V_{j}\oplus V_{j-\frac{1}{2}},
\ee
where  $V_{j}$ stands for the state space of a spin-$j$ representation of $\mf{sl}(2)$ which is $2j+1$ dimensional, thus the direct sum (\ref{repth7}) is $4j+1$ dimensional. Any state of the representation (\ref{repth7}) is characterized by two variables $(l(j), m)$, one can denote any state of  (\ref{repth7}) as follows

\be \label{repth8}
|l(j) ,m\rangle,
\ee
where $l(j)$ denotes the subspace ($V_{j}$ or $V_{j-\frac{1}{2}}$) to which the state (\ref{repth8}) belongs, thus $l(j)$ takes two values 
\be \label{repth8-1}
l(j) = \begin{cases} j, \\ j-1/2, \end{cases}
\ee
$m$ is the $\mf{sl}(2)$ spin projection of $l(j)$, hence
\be  \label{repth9}
\begin{split}
&m \in [-j, -j+1, ..., j-1, j] \qquad \text{if    $l(j)=j$ or,} \\ &
m \in[ -j+\frac{1}{2}, -j+\frac{3}{2}, ..., j-\frac{3}{2}, j -\frac{1}{2}] \qquad \text{if    $l(j)=j-\frac{1}{2}$}.
\end{split}
\ee  
In the sequel we will use the notation
\be  \label{repth9-1}
l_i= l_i(j_i).
\ee

The generators of the $\mf{osp}(1|2)$ algebra in a superspin-$j$ representation will be denoted as follows
\be  \label{repth9-2}
L_{0},L_{\pm 1}, G_{\pm \frac{1}{2}}, 
\ee
with the commutation relations (\ref{repth2})\footnote{To distinguish conformal algebra and CS gauge algebra generators we use bold and regular fonts respectively.}. The generators (\ref{repth9-2}) act on the states\footnote{In the sequel, the superscript $i$ over generators labels superspins-$j_i$ representation.} (\ref{repth8}) as follows \cite{Ennes:1997vt, berezin1981group}
\be \label{repth10}
\begin{split}
& L_0 | l(j) ,m\rangle= m | l(j) ,m\rangle, \\
& L_{\mp 1}  | l(j) ,m\rangle = \mp \sqrt{     [ j\mp m][j \pm m+1] }  | l(j) ,m \pm 1 \rangle, 
\end{split}
\ee
\be \label{repth11}
 G_{\mp \frac{1}{2}}  | \tilde{l}_1(j) ,m\rangle= \begin{cases}  \pm  i \sqrt{j\mp m }  |  \tilde{l}_2(j) ,m \pm \frac{1}{2}\rangle, & \text{if $j-m \in \mathbb{Z}$ } ,   \\   i\sqrt{j\pm m +\frac{1}{2}} | \tilde{l}_2(j) ,m \pm \frac{1}{2}\rangle, & \text{If $j-m\in \mathbb{Z}+\frac{1}{2}$},
 \end{cases} 
\ee
where $[x]$ represents the integer part of the number $x$ ($2x \in \mathbb{Z}$  \footnote{$\mathbb{Z}$ is the set of integers.}), and if $\tilde{l}_1(j)=j$ then $\tilde{l}_2(j)=j-\frac{1}{2}$ or vice-versa.

 The second fact is related to the structure of the Clebsch-Gordan coefficients (CGC) of $\mf{osp}(1|2)$. The tensor product of two representations $(\mathcal{R}_{j_1}, \mathcal{R}_{j_2})$ decomposes in the direct sums
 
 \be \label{repth12}
 \mathcal{R}_{j_1}\otimes \mathcal{R}_{j_2}= \bigoplus_{j_3=|j_1-j_2| \atop 2(j_3-j_1-j_2)\in \mathbb{Z}} ^{j_1+j_2}  \mathcal{R}_{j_3},
 \ee
any state in the tensor product can be expressed as follows
\be \label{repth13}
\big{ |}   l_{3}, m_{3} \big \rangle   = \sum_{l_1, l_2} \sum_{m_1, m_2} C( l_1, m_1;l_2, m_2| l_{3}, m_{3}) \big{|}   l_{1}, m_{1} \big \rangle \otimes \big{|}  l_{2}, m_{2} \big \rangle,
\ee  
where 

\be  \label{repth14}
C( l_1, m_1; l_2, m_2| l_{3}, m_{3})
\ee
are the $\osp$ Clebsch-Gordan coefficients, from (\ref{repth12}) it follows that $j_{3}$ in $l_3=l_3(j_3)$ can take values  
\be  \label{repth15}
j_{3} \in  \Big{[} |j_1-j_2|, |j_1-j_2|+\frac{1}{2},..., j_1+j_2-\frac{1}{2}, j_1+j_2 \Big{]}.
\ee
In \cite{Scheunert:1976wj,  berezin1981group, daumens1993super} has been shown that the CGC (\ref{repth14}) can be factorized as follows
\be \label{repth16}
 C( l_1, m_1; l_2, m_2   | l_{3}, m_{3})=  \begin{bmatrix}  j_1  & j_2& j_3  \\  l_1 & l_2 &l_3 \end{bmatrix} C_{\mf{sl}(2)}\left(l_1, m_1;l_2, m_2|l_{3},m_{3}\right),
\ee 
where $ C_{\mf{sl}(2)}\left(l_1, m_1;l_2, m_2|l_{3},m_{3}\right)$ are the CGC of $\mf{sl}(2)$ (\ref{wl26}) and
\be\label{repth17}
\begin{bmatrix}  j_1  & j_2& j_3  \\  l_1 & l_2 &l_3 \end{bmatrix}
\ee
is termed \textit{the symmetrical scalar factor}. The scalar factor (\ref{repth17}) does not depend on the projections $(m_1, m_2, m_{3})$, it has been computed in \cite{daumens1993super}. 

The singlet state (the state with the total superspin and projection equal to zero) in the tensor product of superspin representations will be denoted as follows 
\be \label{repth18}
| 0,0 \rangle= | s  \rangle.
\ee
In the case of the tensor product of two representations (\ref{repth12}) with the same superspin $j_1=j_2=j$ the CGC of the singlet are given by \cite{daumens1993super} 
\be \label{repth19}
C(l_1, m_1;  l_2, m_2|0,0)=(-1)^{2(\lambda+1)(j-l_1)+l_1-m_1}\delta_{l_1 l_2}\delta_{m_1, -m_2},
\ee
where $\lambda$ is the parity of $j$. 

Similarly to $\mf{sl}(2)$, in $\mf{osp}(1|2)$ it is defined the super $3-j$ symbol given by the following formula \cite{daumens1993super}
\be \label{repth20}
\begin{pmatrix}  j_1  & j_2 & j_3  \\ l_1m_1 & l_2 m_2 & l_3m_3 \end{pmatrix}=\begin{bmatrix}  j_1  & j_2& j_3  \\  l_1 & l_2 &l_3 \end{bmatrix} \begin{pmatrix}  l_1 & l_2 & l_3 \\ m_1 & m_2 & m_3 \end{pmatrix},
\ee
where on the rhs the second factor is the Wigner $3-j$ symbol of $\mf{sl}(2)$ algebra.  Using the super $3-j$ symbol we can express the singlet state of the tensor product of three representations as follows 
\be  \label{repth21}
| s \rangle=     \sum_{l_i, m_i}  (-1)^{l_1-l_2+l_3} \begin{pmatrix}  j_1  & j_2 & j_3  \\ l_1m_1 & l_2 m_2 & l_3m_3 \end{pmatrix}  |  l_1, m_1         \rangle \otimes  |  l_2, m_2         \rangle \otimes  |  l_3, m_3         \rangle,
\ee
where the sums are over all possible values of $l_i, m_i$ according to (\ref{repth8-1}, \ref{repth9}, \ref{repth9-1}). Notice that for fixed values of $(l_1, l_2, l_3)$ in (\ref{repth21}) the sums over $(m_1, m_2, m_3)$  yield a singlet state in the tensor product of three $\mf{sl}(2)$ finite-dimensional representations.

\section{Wilson line construction}  \label{confb}
In this section we will compute using the Wilson lines formulation the CBs of the components of the two-point, three-point correlation functions on the sphere and one-point correlation function on the torus.

\subsection{Two- and three-point spherical conformal blocks} \label{sph}
Here we show that the CBs of each component of the two-point correlation function (\ref{cor3}) and three-point correlation function (\ref{cor6}) can be obtained by an ansatz similar to (\ref{ni3}).   We introduce the following notation
\be \label{wl1}
A_i^k(z) = \begin{cases}   \varphi_i (z)   \quad \text{if $k=0$,} \\ \psi_i(z)  \quad  \text{if $k=1$.}     \end{cases} 
\ee
As discussed in \cite{Bhatta:2016hpz, Bhatta:2018gjb, Besken:2016ooo, Fitzpatrick:2016mtp, Alkalaev:2020yvq} the Wilson line description of CBs implies relations between conformal dimensions $h_i$ and spins $j_i$ (in our case superspins $j_i$) as well as relations between primary fields $\varphi_i$ and states of $\mathcal{R}_{j_i}$, in the case of $\mf{osp}(1|2)$ we find the following relations\footnote{For the other two components of (\ref{ni5}) the relation is $\bar{\psi}_i(z) \rightarrow  |  j_i,  j_i   \rangle$  and  $\tilde{\varphi}_i(z) \rightarrow  |  j_i-\frac{1}{2},  j_i-\frac{1}{2}   \rangle$. } (the arrows stand for corresponds)
\be \label{wl2}
\begin{split}
&h_i=-j_i, \\ & \varphi_i(z) \rightarrow  |  j_i,  j_i   \rangle,  \\  &  \psi_i(z) \rightarrow  |  j_i-\frac{1}{2},  j_i-\frac{1}{2}   \rangle. 
\end{split}
\ee

On the sphere, by taking into account the relations (\ref{wl2}) and using the notations (\ref{wl1}) we will have that the components of the correlation functions (\ref{cor2}, \ref{cor4}) can be expressed by the following formula
\be \label{wl2-1} 
\bigg \langle  \prod_{i=1}^n A_{i}^{k_i} (z_i) \bigg \rangle = c \bigg \langle  s \bigg{ |} \prod_{i=1}^n   e^{\int_{z_i}^{z_b} L_1^{(i)}dz}\bigg{|}j_i-\frac{k_i}{2} , j_i-\frac{k_i}{2}       \bigg  \rangle,
\ee
where, $n=2$ (resp. $3$)  in the case of two-point (resp. three-point) correlation function,  $c$ is a constant irrelevant to our discussions, $ \langle  s  |  $ is a singlet state in the tensor product of superspin representations $\mathcal{R}_{j_i}$ of $\mf{osp}(1|2)$, $L_1^{(i)}$ is the lowering generator acting on the states of $\mathcal{R}_{j_i}$, $| j_i-k_i/2, j_i-k_i/2  \rangle $ (for $k_i=0,1$) is a state in $\mathcal{R}_{j_i}$ according to the notation (\ref{repth8}) and $z_b$ is an arbitrary point (the CBs do not depend on it). To see that indeed (\ref{wl2-1})  does not depend on $z_b$ let  us choose other point $z_{b'}$ in (\ref{wl2-1}) and see that it gives the same result, thus we have
\be
\begin{split}
& \bigg \langle  s \bigg{ |} \prod_{i=1}^n   e^{\int_{z_i}^{z_{b'}} L_1^{(i)}dz}\bigg{|}j_i-\frac{k_i}{2} , j_i-\frac{k_i}{2}       \bigg  \rangle= \bigg \langle  s \bigg{ |}  \left( \prod_{i=1}^n   e^{\int_{z_b}^{z_{b'}} L_1^{(i)}dz} \right)  \prod_{i=1}^n   e^{\int_{z_i}^{z_{b}} L_1^{(i)}dz}\bigg{|}j_i-\frac{k_i}{2} , j_i-\frac{k_i}{2}       \bigg  \rangle=  \\ & = \bigg \langle  s \bigg{ |} \prod_{i=1}^n   e^{\int_{z_i}^{z_{b}} L_1^{(i)}dz}\bigg{|}j_i-\frac{k_i}{2} , j_i-\frac{k_i}{2}       \bigg  \rangle,
\end{split}
\ee
where we used the property of the singlet $\langle  s |$ that  it is annihilated by the action of the element $\hat{L}_1= \sum_{i=1}^n L_1^{(i)} $ because the singlet has total superspin and projection equal to zero in the tensor product, and hence the singlet $\langle  s |$ is invariant under the action of the group element $ \langle s |  e^{(z_{b'}-z_b) \hat{L}_1}=\langle s|$.

\textbf{The two-point correlation function}. By using the relations (\ref{wl2}) and the formula (\ref{wl2-1}) we find that the CBs of the two components (\ref{cor3}) can be computed by
\be \label{wl3}
\langle \varphi_1(z_1)  \varphi_1(z_2)  \rangle \propto  \langle  s | e^{(z_1-z_b)L_1^{(1)}} | j_1, j_1     \rangle \otimes    e^{(z_2-z_b)L_1^{(1)}} | j_1 , j_1  \rangle, 
\ee 

\be \label{wl4}
\langle \psi_1(z_1)  \psi_1(z_2)  \rangle \propto  \langle  s | e^{(z_b-z_1)L_1^{(1)}} | j_1-\frac{1}{2}, j_1-\frac{1}{2}     \rangle \otimes    e^{(z_b-z_2)L_1^{(1)}} |  j_1-\frac{1}{2} , j_1-\frac{1}{2}  \rangle, 
\ee   
where $   \langle   s|$ is the singlet of the tensor product $\mathcal{R}_{j_1}\otimes \mathcal{R}_{j_1}$. Choosing $z_b=z_2$ and considering the coefficients of the singlet (\ref{repth19}), the rhs of (\ref{wl3}) reduces (up to irrelevant constant) to
\be \label{wl5}
\langle   -j_1, j_1 | e^{-z_{12}L_1^{(1)}} | j_1, j_1     \rangle ,
\ee
where $z_{ij}=z_i-z_j$, by applying the generator $L_1^{(1)}$ to the states in $\mathcal{R}_{j_1}$ according to (\ref{repth10}) we obtain
\be \label{wl6}
\langle \varphi_1(z_1)  \varphi_1(z_2)  \rangle \propto  (z_1-z_2)^{2j_1},
\ee
which confirms (\ref{wl3}). Similarly  for (\ref{wl4}) we obtain
\be  \label{wl7}
\langle \psi_1(z_1)  \psi_1(z_2)  \rangle \propto  (z_1-z_2)^{2j_1-1}.
\ee

\textbf{The three-point correlation function}. According to (\ref{wl2-1}) and the singlet (\ref{repth21}), we have in this case the following equation
\be  \label{wl8}
\begin{split}
 &  \langle  A_1^{k_1}(z_1) A_2^{k_2}(z_2)A_3^{k_3}(z_3)  \rangle = C_1(j_i, k_i)   \sum_{l_i, m_i} \Bigg{[} (-1)^{l_1-l_2+l_3} \begin{pmatrix}  j_1  & j_2 & j_3  \\ l_1m_1 & l_2 m_2 & l_3m_3 \end{pmatrix}  \times \\ & \times  
  \Big \langle m_1, l_1 \big{|}  e^{z_{b1}L_1^{(1)}}     \big{|}   j_1-\frac{k_1}{2}, j_1-\frac{k_1}{2} \Big \rangle \times    \Big \langle  m_2, l_2          \big{|}  e^{z_{b2}L_1^{(2)}}   \big{|}   j_2-\frac{k_2}{2}, j_2-\frac{k_2}{2} \Big \rangle  \times      \\ &  \times  \Big \langle  m_3, l_3          \big{|}  e^{z_{b3}L_1^{(3)}}    \big{|}   j_3-\frac{k_3}{2}, j_3-\frac{k_3}{2}     \Big \rangle \Bigg{]},
  \end{split}
\ee
where in this sum $l_i$ takes values of $j_i$ and $j_i-1/2$ , and $m_i$ takes values according to (\ref{repth9}). By taking into account  the expression (\ref{repth20}) of super $3-j$ symbol and since the action of the generator $L_1^{(i)}$ on $|  l_i, m_i  \rangle$ does not change the value of $l_i$, the rhs of the above equation reduces to
\be \label{wl9}
\begin{split}
&C_1(j_i, k_i) (-1)^{j_1-j_2+j_3+\frac{k_1-k_2+k_3}{2}}\begin{bmatrix}  j_1  & j_2 & j_3  \\  j_1-\frac{k_1}{2} & j_2-\frac{k_2}{2}  &j_3-\frac{k_3}{2}  \end{bmatrix} \times   \\ & 
\sum_{m_i} \Bigg{[} \begin{pmatrix}  j_1-\frac{k_1}{2}  & j_2-\frac{k_2}{2} & j_3-\frac{k_3}{2}  \\ m_1 & m_2 & m_3 \end{pmatrix}  
  \Big \langle m_1,  j_1-\frac{k_1}{2}  \big{|}  e^{z_{b1}L_1^{(1)}}     \big{|}   j_1-\frac{k_1}{2} , j_1-\frac{k_1}{2}  \Big \rangle    \times \\ &\times     \Big \langle m_2,  j_2-\frac{k_2}{2}       \big{|}  e^{z_{b2}L_1^{(2)}}   \big{|}   j_2-\frac{k_2}{2}  , j_2-\frac{k_2}{2}   \Big \rangle \times    \Big \langle m_3,  j_3-\frac{k_3}{2}        \big{|}  e^{z_{b3}L_1^{(3)}}    \big{|}   j_3-\frac{k_3}{2} , j_3-\frac{k_3}{2}      \Big \rangle \Bigg{]},
\end{split}
\ee  \label{wl10}
this expression can be shown \cite{Besken:2016ooo} to be proportional to $ z_{12}^{S_1+S_2-S_3}  z_{13}^{S_1+S_3-S_2}  z_{23}^{S_2+S_3-S_1}$, where $S_i= j_i-k_i/2$, thus we obtain
\be  \label{wl11}
 \langle  A_1^{k_1}(z_1) A_2^{k_2}(z_2)A_3^{k_3}(z_3)  \rangle = C(j_i, k_i, \lambda_i) z_{12}^{S_1+S_2-S_3}  z_{13}^{S_1+S_3-S_2}  z_{23}^{S_2+S_3-S_1}, 
\ee
where $C(j_i, k_i, \lambda_i)$ is a constant. Notice that (\ref{wl9}) does not vanish only when $S_1+S_2+S_3$ is an integer that is the condition when the structure constant $C_{123}$ (or $\tilde{C}_{123}$) does not vanish.  

\subsection{One-point torus conformal block} \label{1torus}
In this section we will compute the lower and upper superblocks (\ref{cor9}, \ref{cor10}) for the case where the conformal dimensions $(h,h_1)$ are nonpositive integers or half-integers (these superblocks are associated with finite-dimensional representations). We will denote them by $B_0(h,h_1)_f, B_1(h,h_1)_f$ where the subscript $f$ denotes finite-dimensional representation. As mentioned above the superblocks (\ref{cor9}, \ref{cor10}) have poles when $h$ is a nonpositive integer or nonpositive half-integer, hence we can not obtain $B_0(h,h_1)_f$, $B_1(h,h_1)_f$ by substituting $h$ with integer or half-integer nonpositive values. In order to compute them we substitute $h$ with integer or half-integer nonpositive values into (\ref{cor9}, \ref{cor10}) and subtract the infinite part\footnote{See \cite{Alkalaev:2020yvq} for details.}, the resulting expressions correspond to the superblocks associated with finite-dimensional representations, thus these superblocks are computed as follows
\be  \label{wl12}
B_0(h=-j, h_1=-j_1,q)_f= B_0(-j, -j_1,q)-B_0(-(j+\frac{1}{2}), -j_1,q),
\ee

\be   \label{wl13}
B_1(h=-j, h_1=-j_1,q)_f= B_1(-j, -j_1,q)-B_1(-(j+\frac{1}{2}), -j_1,q).
\ee
By computing these differences and taking into account only the first $4j+1$ terms\footnote{Because there are only $4j+1$ states in finite-dimensional representations.} in the expansion in $q$ we obtain the following expressions

\be   \label{wl14}
B_0(-j, -j_1,q) _f= \cF(-j, -j_1,q)-\frac{2j-j_1}{2j}\cF(-(j-1/2), -j_1,q),
\ee

\be   \label{wl15}
B_1(-j, -j_1,q)_f = \cF(-j, -j_1+\frac{1}{2},q)-\frac{2j+j_1+\frac{1}{2}}{2j}\cF(-(j-1/2), -j_1+\frac{1}{2},q),
\ee
where $\cF(-j, -j_1,q)$ is  the $\mf{sl}(2)$ one-point torus block given by (\ref{aa2}).

Now, we want to prove that superblocks (\ref{wl14}, \ref{wl15}) can be computed  by the following expression (up to an irrelevant constant $c(j,j_1,k)$)
\be  \label{wl18}
\text{\Large{Tr}}_j\Bigg{(}W_j [z_b, z_b+2\pi \tau] I_{j;j, j_1}   \Bigg{)}\otimes W_{j_1}[z_b, z_1]|  j_1-\frac{k}{2}, -j_1+\frac{k}{2}  \rangle =c(j,j_1,k) B_k(-j, -j_1,q)_f,
\ee
where $k=0,1$. $|  j_1-\frac{k}{2},- j_1+\frac{k}{2} \rangle$ is a state of the superspin  representation $\mathcal{R}_{j_1}$ of $\mf{osp}(1|2)$, Tr$_j$ denotes the trace over the supermodule $\cV_j$ (\ref{repth7}) of the superspin representation $\mathcal{R}_j$, $z_b$ is an arbitrary point in the bulk,  $z_1$ is a point corresponding to the position of the fields $(\varphi_1, \psi_1)$ at the boundary\footnote{The boundary corresponds to $\rho=\infty$ (for details, see \cite{Alkalaev:2016ptm, Alkalaev:2017bzx, RamosCabezas:2020mew}).} (the one-point blocks do not depend on $z_1$). $W_a[z_1, z_2]$ are the Wilson line operators 
\be 
W_{a}[z_1,z_2]= \exp \left( (z_1-z_2) (   L_1+\frac{1}{4} L_{-1}  ) \right), \,
\ee
here $L_1$ and $L_{-1}$ are the generators of the $\mf{osp}(1|2)$ gauge algebra which act on the states of the respective representation $\mathcal{R}_a$. $I_{j;j,j_1}$ in (\ref{wl18}) is the intertwining operator. In general one can have the intertwining operator $I_{j_3;j_1,j_2}$ associated with representations $(\mathcal{R}_{j_1}, \mathcal{R}_{j_2}, \mathcal{R}_{j_3})$, it acts as follows 
\be \label{wl19}
I_{j_3;j_1,j_2}: \qquad \mathcal{R}_{j_1}\otimes \mathcal{R}_{j_2}\rightarrow \mathcal{R}_{j_3},
\ee 
and satisfies the following defining condition
\be  \label{wl20}
I_{j_3;j_1,j_2}U_{j_1}U_{j_2}= U_{j_3}I_{j_3;j_1,j_2},
\ee
where $U_{j_i} $ (for $i=1,2,3$) are the elements of the $OSP(1|2)$ group in representations $\mathcal{R}_{j_i}$. By taking into account these defining properties of the intertwining operator we have the following matrix elements in terms of the CGC
\be   \label{wl21}
\bigg{\langle}   m_{3}, l_{3} \bigg{|}I_{j_3;j_1,j_2} \bigg {|}    l_{1}, m_{1} \bigg \rangle \otimes \bigg{|} l_{2}, m_{2} \bigg \rangle =  C( l_1, m_1; l_2, m_2| l_{3}, m_{3}),
\ee  
where the CGC are given by (\ref{repth16}). The proof of the relation (\ref{wl18}) is based on the structure (\ref{repth16}) of the CGC and the Wilson lines formulation of the $\mf{sl}(2)$ torus block (see \cite{Alkalaev:2020yvq}).    

We decompose the trace $Tr_j$ in (\ref{wl18}) into two subtraces according to (\ref{repth7}), thus the lhs of (\ref{wl18}) becomes
\be  \label{wl22}
\begin{split}
&\sum _{m\in M_1} \bigg \langle     m, j \bigg{|} W_j [z_b, z_b+2\pi \tau] I_{j;j, j_1}\bigg{|}      j, m  \bigg \rangle \otimes W_{j_1}[z_b, z_1]\big{|}  j_1-\frac{k}{2}, -j_1+\frac{k}{2} \big \rangle  + \\ &
+\sum _{m\in M_2} \bigg \langle     m, j-\frac{1}{2} \bigg{|} W_j [z_b, z_b+2\pi \tau] I_{j;j, j_1}  \bigg{|}      j-\frac{1}{2} , m \bigg \rangle \otimes W_{j_1}[z_b, z_1]   \big{|}  j_1-\frac{k}{2}, -j_1+\frac{k}{2} \big \rangle  ,
\end{split}
\ee
where 
\be   \label{wl23}
M_1 = \Big{[} -j,- j+1,..., j-1, j\Big{]},
\ee

\be    \label{wl24}
M_2 = \Big{[} -j+1/2,- j+3/2,..., j-3/2, j-1/2 \Big{]}.
\ee
By taking into account (\ref{repth16}, \ref{wl21}) and that the action of the operator $L_1$ of the Wilson line operators $W_j, W_{j_1}$ on the states $| l_i, m_i \rangle$ do not change the  values of $l_i$ (see \ref{repth10}), we can rewrite (\ref{wl22}) as follows
\be  \label{wl25}
\begin{split}
&\mathbf{F}_1 \sum _{m\in M_1} \bigg \langle     m, j \bigg{|} W_j [z_b, z_b+2\pi \tau] I^{\mf{sl}_2}_{j;j, j_1}  \bigg{|}      j, m  \bigg \rangle \otimes W_{j_1}[z_b, z_1]\big{|}  j_1-\frac{k}{2}, -j_1+\frac{k}{2} \big \rangle  + \\ &
+\mathbf{F}_2\sum _{m\in M_2} \bigg \langle     m, j-\frac{1}{2} \bigg{|} W_j [z_b, z_b+2\pi \tau] I^{\mf{sl}_2}_{j;j, j_1}  \bigg{|}      j-\frac{1}{2} , m \bigg \rangle  \otimes W_{j_1}[z_b, z_1]\big{|}  j_1-\frac{k}{2}, -j_1+\frac{k}{2} \big \rangle ,
\end{split}
\ee
where $(\mathbf{F}_1 , \mathbf{F}_2)$ are two factors (\ref{repth17})
\be \label{wl25-1}
\begin{split}
&\mathbf{F}_1 =    \begin{bmatrix} 
   j &j _1 &j  \\j &  j_1-\frac{k}{2} & j \end{bmatrix} =(2j+j_1+1-k/2)^{\frac{1}{2}}, \\ 
&\mathbf{F}_2 = \begin{bmatrix} j&j_1  &j \\  j-\frac{1}{2}  &   j_1-\frac{k}{2}    & j-\frac{1}{2}  \end{bmatrix} =(-1)^{\lambda+1}(2j-j_1+k/2)^{\frac{1}{2}},
\end{split}
\ee
and $I^{\mf{sl}_2}_{j_3;j_1, j_2}$ in (\ref{wl25}) acts as follows (for fixed values of $l_i$)  
\be    \label{wl26}
\begin{split}
 \langle   l_{3}, m_{3} | I^{\mf{sl}_2}_{j_3;j_1, j_2} |  l_{1}, m_{1}  \rangle \otimes |   l_{2}, m_{2}  \rangle &=  C_{\mf{sl}(2)}\left(l_1, m_1;l_2, m_2|l_{3},m_{3}\right)    =  \\ &=(-1)^{-j_1+j_2-m_3}\sqrt{2j_3+1}   \begin{pmatrix}  j_1 & j_2 & j_3 \\ m_1 & m_2 & -m_3 \end{pmatrix}.
 \end{split}
\ee  
As the spherical CBs (\ref{wl25}) does not depend on $z_b$ and due to the translational invariance of the one-point torus block we can choose $z_1=z_b=0$. In order to simplify the computation of (\ref{wl25}) we diagonalize $W_j [z_b, z_b+2\pi \tau] =e^{2\pi \tau (L_1+\frac{1}{4}L_{-1})}$ expressing it in the following way
\be \label{wl26-1}
W_j [z_b, z_b+2\pi \tau] = U_j  e^{2\pi i \tau L_0}U_j^{-1}=U_j  q^{L_0}U_j^{-1},
\ee
where $U_j= e^{\frac{i}{2}L_{-1}}e^{-iL_1}e^{-i \pi L_0}$ is a group element in $\mathcal{R}_j$. Placing (\ref{wl26-1}) in (\ref{wl25}) and   due  to the property (\ref{wl20}) of the intertwining operator, after this replacement, (\ref{wl25}) remains unchanged except the boundary state $| j_1-\frac{k}{2}, -j_1+\frac{k}{2} \rangle$ which transforms 
\be  \label{wl26-2}
| j_1-\frac{k}{2}, -j_1+\frac{k}{2} \rangle  \rightarrow U_{j_1}^{-1}| j_1-\frac{k}{2}, -j_1+\frac{k}{2} \rangle=\sum_{s=-j_1+k/2}^{j_1-k/2}\beta_s| j_1-\frac{k}{2}, s \rangle , 
\ee
where $\beta_s$ are some coefficients depending on ($j_1, k,s$), in fact, we will need only the state with projection $s=0$. By  taking into account (\ref{repth10}, \ref{wl26},  \ref{wl26-1}, \ref{wl26-2}) we have that (\ref{wl25}) becomes
\be \label{wl26-3}
\begin{split}
& {}  \mathbf{F}_1 \sum _{m\in M_1} \bigg \langle     m, j \bigg{|} q^{L_0}I^{\mf{sl}_2}_{j;j, j_1}  \bigg{|}      j, m  \bigg \rangle \otimes U_{j_1}^{-1}  {|}  j_1-\frac{k}{2}, -j_1+\frac{k}{2} \big \rangle  + \\ + &
\mathbf{F}_2\sum _{m\in M_2} \bigg \langle     m, j-\frac{1}{2} \bigg{|}  q^{L_0}   I^{\mf{sl}_2}_{j;j, j_1}  \bigg{|}      j-\frac{1}{2} , m \bigg \rangle  \otimes U_{j_1}^{-1}  \big{|}  j_1-\frac{k}{2}, -j_1+\frac{k}{2} \big \rangle= \\ &=  \mathbf{F}_1   \sum _{m\in M_1,s}  q^m (-1)^{-m-j+j_1-k/2}\sqrt{2j+1}   \begin{pmatrix}  j& j_1-\frac{k}{2} &  j  \\ m & s& -m \end{pmatrix}\beta_s  +\\ &+ \mathbf{F}_2  \sum _{m\in M_2,s}  q^m (-1)^{-m-j+ \frac{1}{2}+  j_1-k/2}\sqrt{2j}  \begin{pmatrix}  j-\frac{1}{2}& j_1-\frac{k}{2} &  j -\frac{1}{2} \\ m & s& -m \end{pmatrix}\beta_s.
\end{split}
\ee
Due to the property of Wigner $3-j$ symbol the nonzero terms correspond to\footnote{The zero projection $s=0$ only exists when $j_1-k/2$ is integer.}  $s=0$, by taking into account this and collecting a common factor $\tilde{c}$ we have (\ref{wl26-3}) is 
\be \label{wl26-4}
\tilde{c}   \bigg{[} \sum _{m\in M_1}  q^m (-1)^{m}   \begin{pmatrix}  j& j_1-\frac{k}{2} &  j  \\ m & 0 & -m \end{pmatrix}  +      \frac {\mathbf{F}_2 (-1)^{\frac{1}{2}} \sqrt{2j}}{\mathbf{F}_1 \sqrt{2j+1}} \sum _{m\in M_2}  q^m (-1)^{m} \begin{pmatrix}  j-\frac{1}{2}& j_1-\frac{k}{2} &  j -\frac{1}{2} \\ m & 0& -m \end{pmatrix} \bigg{]}  .
\ee
Each sum in (\ref{wl26-4}) can be shown \cite{Alkalaev:2020yvq} to be proportional to $\mf{sl}(2)$ one-point torus block for finite-dimensional representations (\ref{aa2}) denoted by $\cF$. By computing\footnote{In (\ref{wl27}) we set $j$ to be an integer.} these sums, (\ref{wl26-4}) can be expressed as follows.

\be \label{wl27}
c(j,j_1,k) \left(  \cF(-j, -j_1+k/2,q)-\frac{2j-(-1)^kj_1+\frac{k}{2}}{2j}\cF(-(j-1/2), -j_1+k/2,q)  \right).
\ee
which proves (\ref{wl18}). We notice that when in (\ref{wl26-3})  there is no such projection $s=0$ then (\ref{wl26-3})  vanishes, this is consistent with the fact that in this case the structure constant $C_{j (j_1-\frac{k}{2})j}$ (\ref{cor8}, \ref{cor8-1}) also vanishes.

\section{Conclusions} \label{conclusion}
In this work we studied the Wilson lines formulation of the $\mf{osp}(1|2)$ CBs. We considered CBs associated with finite-dimensional representations and light $\Phi_i$ primary superfields. We used the decomposition of superfields into ordinary primary fields (\ref{ni5})   and we concentrated only on the $(\varphi_i, \psi_i)$ sectors, thus we expressed the correlation functions of superfields in terms of ordinary fields, (eqs. (\ref{ni7}, \ref{ni8})). On the sphere, we showed that the CBs of the components of the two and three-point correlation functions can be computed by the Wilson lines formulation (\ref{wl2-1}). On the torus, we computed the lower and upper superblocks (\ref{wl14}, \ref{wl15}) via the Wilson lines formulation (\ref{wl18}). The ansatz (\ref{wl2-1}, \ref{wl18}) involve three main components: the singlet states belonging to the tensor product of superspin$-j$ representations of $\mf{osp}(1|2)$; Wilson line operator built out of the elements of $\mf{osp}(1|2)$ algebra; boundary states of superspin$-j$ representations of $\mf{osp}(1|2)$. The ansatz (\ref{wl2-1}, \ref{wl18}) is very similar to the ansatz of the ordinary Virasoro case, and its proof essentially reduces to the $\mf{sl}_2$ case. This reduction is related to the facts that the state space (\ref{repth7}) of a superspin representation of $\mf{osp}(1|2)$ splits into two normal subspaces of two spin representations of $\mf{sl}_2$ and that the CGC of $\mf{osp}(1|2)$ factorizes into the CGC of the $\mf{sl}_2$ and the scalar factor as in (\ref{repth16}). We also found in the super-Virasoro case that different components ($\varphi_i, \psi_i$) of the superfield $\Phi_i$  are related to different states of a superspin representation of $\mf{osp}(1|2)$ according to (\ref{ni12}, \ref{wl2}), which make the dual construction more general.

According to \cite{Henneaux:1999ib} there are seven classes of superalgebras that involve extra internal degrees of freedom and realize the AdS$_3$ gravity in the CS formulation. In this work we have covered one of those cases, the $\mf{osp}(1|2)$. It might be interesting to extend the explicit construction of CBs in the Wilson lines context to the remaining classes. Another natural generalization consists in formulating the dual construction in the full $N=1$ NS super-Virasoro CFT. In particular, this requires the study of quantum corrections (for related consideration, see \cite{Hikida:2018dxe}). It is also interesting to consider CBs of $\mf{sl}_3$ algebra in higher genus topologies. $\mf{sl}_3$ algebra is relevant for $W_3$ CFT in the large central limit. This consideration might be useful for a better understanding of CBs which are not completely fixed by the symmetry algebra in $W_3$ CFT (see, e.g. \cite{Belavin:2016qaa}). 
\newline
\newline
\textbf{Acknowledgments}. J.R.C. is grateful to the organizers of the Workshop on integrability held at the Leibniz University Hannover and the Israeli Physical Society conference 2022 held at the Ben-Gurion University for giving the opportunity to participate and report this work.

\appendix
\section{$N=1$ superconformal field theory} \label{scft}
The $N=1$ supersymmetric CFT$_2$ is generated by the holomorphic and antiholomorphic components of the stress-energy tensor $T$ and the spin-vector current $J$. The operator product expansion has the following form:
\be \label{N1}
\begin{split}
& T(z)T(w) = (3c/4)/(z-w)^4+\frac{2}{(z-w)^2}T(w)+\frac{1}{z-w}T'(w)+...,\\
&T(z)J(w)= \frac{3}{2}(z-w)^{-2}J(w)+(z-w)^{-1}J'(w)+..., \\ &
J(z)J(w)= c(z-w)^{-3}+(z-w)^{-1}T(w)+...\quad .
\end{split}
\ee 
Equations (\ref{N1}) imply that the Laurent components of $T$ and $J$ satisfy 

\be \label{N2}
\begin{split}
&[\mathbf{L}_n, \mathbf{L}_m]= (n-m)\mathbf{L}_{m+n}+\frac{c}{8}(n^3-n)\delta_{n, -m}, \\
& \{  \mathbf{G}_r, \mathbf{G}_s \}= 2\mathbf{L}_{r+s}+\frac{1}{2}c(r^2-\frac{1}{4})\delta_{r,-s}, \\
&[\mathbf{L}_n, \mathbf{G}_r]= (\frac{n}{2}-r)\mathbf{G}_{n+r},
\end{split}
\ee
where $(n,m)$ take integer values and in the Neveu-Schwarz sector
\be \label{N3}
r,s\in \mathbb{Z}+1/2.
\ee
The hwv $|h_i\rangle $ is generated by the action of the superfield (\ref{ni5}) on the vacuum $| 0 \rangle$
\be \label{N4}
|h_i \rangle=\Phi_i(0)| 0\rangle. 
\ee
the action of $(\mathbf{L}_n$, $\mathbf{G}_r)$ (for positive $n,r$) on the hwv (\ref{N4}) vanishes, the other states  of the supermodule are obtained by applying the operators ($\mathbf{L}_{-n}$, $\mathbf{G}_{-r}$) on the hwv. 

The inner product of a state of the supermodule can be computed using (\ref{N2}), for example for positive integers $(n,r)$
\be \label{N5}
\begin{split}
& \langle h_i |  \mathbf{L}_n  \mathbf{L}_{-n}| h_i \rangle = (2nh_i    +\frac{c}{8}(n^3-n))\langle h_i | h_i \rangle, \\ &
\langle h_i |  \mathbf{G}_r  \mathbf{G}_{-r}| h_i \rangle =(2h_i    +\frac{c}{2}(r^2-\frac{1}{4}))\langle h_i | h_i \rangle.
\end{split}
\ee
In order to have finite inner products (\ref{N5})  in the limit $c \rightarrow \infty$ we need to restrict the values of ($n,r$) to ($n=-1,0,1$; $r=-1/2, 1/2$), this restriction implies that (\ref{N2}) becomes the $\mf{osp}(1|2)$ algebra (\ref{repth2}).

\section{$\mf{sl}(2)$ one-point torus block} \label{sl2b}
The $\mf{sl}(2)$ one-point torus block, for general intermediate and external conformal dimensions ($\tilde{\Delta}, \Delta$) respectively, can be written in terms of the hypergeometric  function as \cite{Hadasz:2009db}
\be
\label{aa1}
\begin{aligned}
&\cF(\tilde\Delta,\Delta, q) = \frac{\;\;q^{ \tilde \Delta}}{1-q} \,\;{}_2 F_{1}(\Delta, 1-\Delta, 2\tilde \Delta\, |\, \frac{q}{q-1})  =
\\
& = 1+\Big[1+\frac{(\Delta - 1)\Delta}{2\tilde \Delta}\Big] q + \Big[1+\frac{(\Delta -1) \Delta }{2 \tilde\Delta }+\frac{(\Delta -1)
 \Delta  (\Delta ^2-\Delta +4 \tilde\Delta)}{4 \tilde \Delta  (2 \tilde\Delta +1)}\Big]q^2+ ... \;.
 \end{aligned}  
\ee
\hfill\break
By using a reasoning similar to the used one in subsection \ref{1torus}, the $\mf{sl}(2)$ one-point torus block associated with nonpositive integer external dimensions $-j_1$ and nonpositive integer intermediate dimensions $-j$ can be extracted from (\ref{aa1}), this CB is given by \cite{Alkalaev:2020yvq}  

\be
\label{aa2}
\cF(-j, -j_1,q)=q^{-j}\left(f_{0} + q  f_{1} + q^2 f_{2} +... + q^{2j} f_{2j}\right),
\qquad
\ee
where $(j, j_1)$ are nonnegative integers or half-integers, and coefficients $f_n$ are given by
\be
\label{a9}
f_{n}  = \, _3F_2(-j_1,j_1+1,-n;1,-2 j;1)= 
\sum_{m=0}^n
\,\frac{(-)^m n!}{(n-m)!(m!)^2}\,\frac{(j_1+m)_{2m}}{(2j)_m},
\ee
where $(x)_n = x(x-1)... (x-n+1)$.

\bibliographystyle{JHEP}
\bibliography{refs}


\end{document}